\documentstyle[emulateapj,onecolfloat,epsf]{article}

\input epsf

\begin{document}
\twocolumn[
\title{Constraints from the Damping Tail}
\author{Martin White}
\affil{Harvard-Smithsonian CfA, 60 Garden St, Cambridge, MA 02138}
\begin{abstract}
The detection of anisotropy in the cosmic microwave background on arcminute
scales by the Cosmic Background Imager (CBI) provides us with our first
measurement of the damping tail and closes one chapter in the CMB story.
We now have experimental verification for all of the features in the
temperature anisotropy spectrum predicted theoretically two decades ago.
The CBI result allows us to constrain both parameterized models based on
the inflationary cold dark matter (CDM) paradigm and to examine model
independent constraints on the matter content, the distance to last
scattering and the thickness of the last scattering surface.
In particular we show that recombination had to proceed ``slowly'', with
the surface of last scattering having a width $\Delta z\ga 50$.  This
provides strong constraints on non-standard recombination scenarios.
By providing a lower limit on the duration of recombination it implies a
lower limit on the polarization of the sub-degree scale anisotropy which
is close to current experimental upper limits.
\end{abstract}
]
\section{Introduction}

Recently the CBI team announced the first detection of anisotropy in the
cosmic microwave background (CMB) on angular scales of several arcminutes
(Padin et al.~\cite{CBI}).
They quote two ``band-powers'', with amplitudes\newline
$\sqrt{\ell(\ell+1)C_\ell/(2\pi)}=59^{+7.7}_{-6.3}\,\mu$K and
$29.7^{+4.8}_{-4.2}\,\mu$K, corresponding to window functions centered at
$\ell=603^{+180}_{-166}$ and $1190^{+261}_{-224}$ respectively.
(Here $\ell$ is the spherical harmonic index in a multipole expansion of
the temperature fluctuation on the sky.)
This detection is important not only for the specific constraints that
it places on theories of the anisotropy spectrum, to which we shall return
shortly, but also because it closes one phase of anisotropy research.

Theoretical models of CMB anisotropy, based on the growth of structure through
gravitational instability in a dark matter dominated universe, predict that
the angular power spectrum of the temperature anisotropy should have 3
distinct parts, separated by two important physical scales
(see e.g.~Bond~\cite{Bond}).
With the 1992 discovery of anisotropy by the {\sl COBE\/} experiment
(Smoot et al.~\cite{COBE}) we obtained experimental verification of the
first part: the plateau in the spectrum at large angular scales (low-$\ell$)
which is generated as photons lose energy climbing out of potentials on
the last scattering surface (Sachs \& Wolfe~\cite{SacWol}).
Numerous experiments (most recently~Miller et al.~\cite{TOCO};
de Bernardis et al.~\cite{Boom}; Hanany et al.~\cite{MAXIMA};
see Tegmark, Zaldarriaga \& Hamilton~\cite{TegZalHam} for a more
complete list)
have now reported detections of an acoustic peak in the power spectrum on
degree scales ($\ell\sim 200$).
The second part of the anisotropy spectrum, this provides us with a snapshot
of sound waves ``in'' the surface of last scattering
(Peebles \& Yu~\cite{PeeYu}; Doroshkevich, Zel'dovich \& Sunyaev~\cite{DZS})
and encodes a wealth of information about cosmology and our model for
structure formation.
Now the detection of fluctuations on scales of several arcminutes
($\ell\sim 10^3$) provides for the first time experimental verification of
final piece of the spectrum: the diffusion damping tail (Silk~\cite{Sil}).
Though the important, predicted polarization is yet to be detected, and much
work remains to refine our knowledge of the spectrum, the current experimental
situation is in remarkable agreement with theoretical predictions made two
decades ago
(Wilson \& Silk~\cite{WilSil}; Silk \& Wilson~\cite{SilWil};
Vittorio \& Silk~\cite{VitSil};
Bond \& Efstathiou~\cite{BonEfs84}, \cite{BonEfs87}),
enhancing our faith in our paradigm for structure formation in the universe.

For many years theorists have been describing what we may learn from ``future''
measurements of the damping tail.
The CBI measurement provides us with the opportunity to finally begin to
implement these claims.
In the following we do so, identifying some constraints on cosmological models
and models for structure formation arising from the CBI data.
Padin et al.~(\cite{CBI}) have already stated limits on a subset of the
popular theoretical models of structure formation, here we point out some
additional constraints on the general paradigm.
These will strengthen considerably as CBI reports data covering more sky and
binned more finely in $\ell$.

\section{The big picture}

Another detection on smaller angular scales ($\ell\sim 5600$;
Dawson et al.~\cite{BIMA}) at a lower amplitude is consistent with
Sunyaev-Zel'dovich (\cite{SZ80}; SZ) fluctuations expected in popular models
of structure formation.
These models predict that the SZ effect is the dominant secondary anisotropy
on arcminute scales, that the bulk of the signal comes from clusters of
galaxies spread over a range of redshifts and that the signal falls off
rapidly toward the larger angular scales probed by CBI.
The CBI detection is of sufficient amplitude that it is highly unlikely to be
secondary anisotropy -- we shall assume from now on that CBI is constraining
the primoridal anisotropy from last scattering.

The CBI measurement shows that the spectrum has begun to damp significantly
by $\ell\sim 1000$, as expected from photon diffusion during recombination
(Silk~\cite{Sil}).
A finer binning in $\ell$ would be required to verify that the damping is
(close to) exponential and to constrain the damping scale, $\ell_D$, more
precisely from the shape of the decline in power.

Within the standard cosmological model, on the scales probed by the CBI
experiment, the CMB power spectrum depends mainly on the primordial power
spectrum, the physical matter density
($\rho_{\rm mat}\propto\omega_{\rm mat}\equiv\Omega_{\rm mat}h^2$),
the baryon density ($\rho_b\propto\omega_b\equiv\Omega_bh^2$) and the
(comoving) angular diameter distance to last scattering $r_\theta$.
(There is a small correction to the spectrum arising from gravitational
lensing which we can safely ignore.)
If we assume that the primordial power spectrum has no sharp features we
can isolate the other three important ingredients of the model:
$\omega_{\rm mat}$, $\omega_b$ and $r_\theta$.

The new observational limit here is on the damping scale, which has the
nice property that it depends primarily on the background
cosmology\footnote{See e.g.~Hu \& White (\protect\cite{Damp}) for a discussion
of technical caveats.}, and not on the assumed model of structure formation
(inflation, defects...).
To a first approximation the damping scale is the geometric mean of the
horizon and the photon mean-free-path just before recombination
(Kaiser~\cite{Kai83}; see Hu \& White~\cite{Damp} for numerical fitting
functions).
Thus an increase in the matter density, which decreases the size of the
horizon at last scattering, will shift the damping to smaller angular scales.
For baryon densities consistent with big-bang nucleosynthesis the damping
scale is also shifted to smaller angular scales by a decrease in the
mean-free-path (an increase in the baryon density).

Let us begin by considering general inferences drawn from the locations of
the gross features (peaks and damping) in the spectrum.
The first acoustic peak appears to lie at $\ell_A\sim200$
(Miller et al.~\cite{TOCO};
de Bernardis et al.~\cite{Boom};
Hanany et al.~\cite{MAXIMA})
providing us with a first rough measurement $\ell_D/\ell_A\sim 5$.
This ratio can be interpreted as the $Q$ of the sound ``cavity''
(fluid at last scattering) and $Q\sim 5$ is in accord with our theoretical
expectations (Hu \& White~\cite{Sig}).
Theoretically $\ell_D/\ell_A$ is independent of distance to last scattering
and only weakly dependent on the assumed energy content,
e.g.~$\rho_{\rm rad}/\rho_\gamma$, and baryon content\footnote{Since $\ell_A$
is independent of $\omega_b$ near $\omega_b=0.02$, the ratio increases slowly
with increasing $\omega_b$.  Current constraints on $\omega_b$
(O'Meara et al.~\cite{BBN}) make this increase negligible.}.
A ratio in the range 4--6 is a strong indication that the fluctuations are
adiabatic, such as are produced `uniquely' by inflation
(Hu, Turner \& Weinberg~\cite{HuTurWei};
 Liddle~\cite{Lid}).
Unfortunately within the current uncertainties on $\ell_D$ the constraints,
while disfavouring a shift in $\ell_D/\ell_A$ by a factor of 1.5 as predicted
by isocurvature models, are not very tight.

Narrowing our attention to adiabatic models, if we combine data on the first
peak with the highest $\ell$ constraint from CBI we can impose a lower limit
on the matter density.  Recall from above that the damping scale is moved to
smaller angular scales as $\omega_{\rm mat}$ is increased.  Holding $\omega_B$
fixed at 0.02 the ratio of the amplitudes of the fourth to the first peak
grows by a factor of 2 as we increase $\omega_{\rm mat}$ from 0.05 to 0.25.
Though this ratio is affected by changes in the spectral index, it is not
affected by changes in the normalization or a late epoch of reionization.
Taking the bandpowers as measures of the power at their central $\ell$, we
estimate that observationally this ratio is close to $1/4$.  This corresponds
to $\omega_{\rm mat}\sim 0.15$ under our assumptions.
Values of $\omega_{\rm mat}\sim\omega_b$ would have this ratio considerably
lower, though experimental uncertainties are such that models of this type
are not (yet) completely ruled out by the current CMB data.
A full study of parameter space and accounting for experimental uncertainties
would be necessary to draw firm conclusions.  Such a study will be highly
informative when additional data from CBI are released, particularly to
higher $\ell$.

\begin{figure}[tbh]
\leavevmode
\centerline{\epsfxsize=3.5in \epsfbox{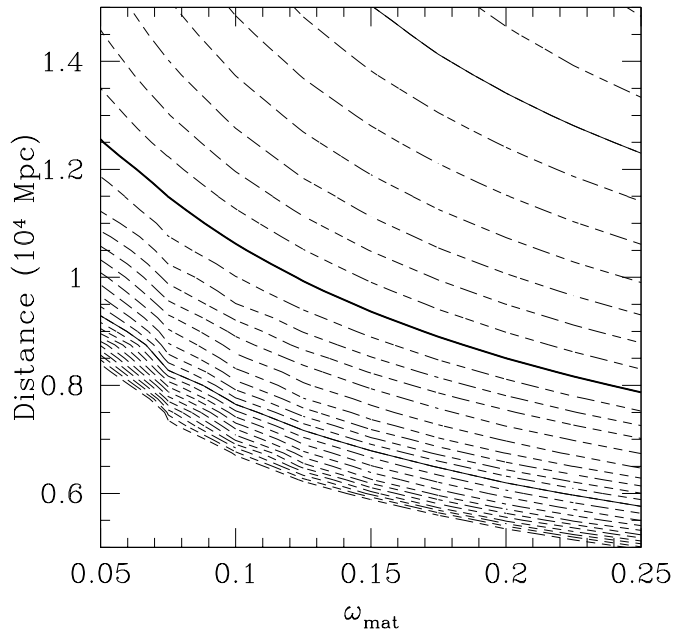}}
\caption{\footnotesize Contours of the bandpower ratio in the
$\omega_{\rm mat}-r_\theta$ plane.  We have chosen $h=0.65$ in computing
distances.  Contours are spaced in steps of 0.1, increasing to the lower
left.  The thick solid line indicates 2, with the thin solid lines being
1.3 and 3.0.}
\label{fig:distance}
\end{figure}

\section{Constraints on models}

Let us now turn to more quantitative constraints on our models.
With only two bandpowers we cannot provide limits on a large parameter
space, so we shall here consider various effects in turn rather than
varying all of them together.
To avoid questions of the calibration uncertainty, the normalization of
the spectrum and any late ($z\la 10^2$) epoch of reionization we shall
focus our attention on the {\it ratio\/} of the bandpowers.
Experimentally this ratio is close to 2, and should lie between
1.3 and 3 at `$2\sigma$'.
When computing theoretical predictions for this ratio we shall approximate
the window functions as Gaussians centered on $\ell=603$ and $1190$ with
$\sigma=104$ and $146$ respectively.
Pearson et al.~(\cite{Pearson}) have shown a Gaussian is a good approximation
for the visibility window function, we shall assume this holds for the
bandpower window function also.

The first limit we shall consider is on the (comoving) angular diameter
distance to last scattering, which is a sensitive function of spatial
curvature:
\begin{equation}
  r_\theta = \left| K \right|^{-1/2} \sinh\left[
  \left| K \right|^{1/2} \left( \eta_0-\eta_* \right) \right]
\end{equation}
for $K<0$ (for positive curvature replace $\sinh$ with $\sin$).
Here $K=H_0^2(\Omega_{\rm tot}-1)$ is the spatial curvature,
$\eta_0=\int dt/a$ is the conformal age of the universe today,
$\eta_*$ is the (conformal) age at last scattering and we have set $c=1$.
At fixed $\omega_{\rm mat}$ and $\omega_b$ the $\ell$ of any feature in
the spectrum depends linearly on this distance.
The amount of power near the damping tail depends exponentially on
$\ell$, so this allows another test of curvature of the universe.

For fixed distance, $r_\theta$, and matter density, $\omega_{\rm mat}$, the
bandpower ratio is quite insensitive to $\omega_b$ so we shall here hold it
fixed at $0.02$ (O'Meara et al.~\cite{BBN}).
We shall also assume that the underlying spectrum is scale-invariant for
simplicity.  Since the bandpowers are separated by a factor of $2$ in angular
scale, a deviation of $\pm 0.2$ in the spectral index translates roughly
into a $\pm 15\%$ change in the bandpower ratio, which can be safely ignored.
Our assumptions can be relaxed when bandpowers covering a wider range of
$\ell$ become available.

Fig.~\ref{fig:distance} shows contours of the bandpower ratio in the
$\omega_{\rm mat}-r_\theta$ plane.
Any upper limit to the distance is quite sensitive to our assumed lower
limit on the bandpower ratio.
The lower limit on the distance is however reasonably robust: for any
reasonable cosmological parameters $r_\theta\ga 6000$Mpc (comoving,
with $h=0.65$).
While this limit could be derived from considering e.g.~the first peak in
the spectrum, the damping tail has the advantage of being less dependent on
the assumption of a particular model for the calculation.

\begin{figure}[tbh]
\leavevmode
\centerline{\epsfxsize=3.5in \epsfbox{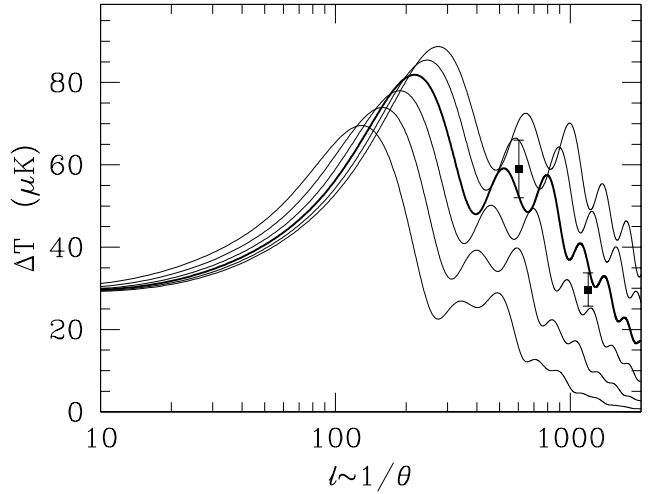}}
\caption{\footnotesize Contribution to the temperature fluctuation per
$\log\ell$ for the concordance model if recombination is modeled as a Fermi
function of width $\Delta z=80$.  From left to right at the first peak,
$z_m=600$ to 1600 in steps of 200.  The thick solid line has $z_m=1200$.
The $y$-axis is $\Delta T\equiv\sqrt{\ell(\ell+1)C_\ell/(2\pi)}$.
The two solid squares are the CBI bandpowers with $\pm 1\sigma$ errors.}
\label{fig:zls}
\end{figure}

Our limit can be interpreted either as a constraint on late-time physics which
changes the distance-redshift relation while holding the redshift of last
scattering (roughly) constant, or on more speculative physics which modifies
the redshift of recombination through e.g.~energy injection.
The first case has been considered by Padin et al.~(\cite{CBI}).  To
illustrate how the second may now be strongly constrained we have calculated
the anisotropies expected for the Ostriker \& Steinhart~(\cite{OstSte})
``concordance model'' --- a standard $\Lambda$CDM model with
$\Omega_{\rm mat}=0.3$, $h=0.67$ and $n=1$ --- replacing the hydrogen ionized
fraction $x_e$ by a Fermi function
$\left( 1+e^{-s} \right)^{-1}$ where $s\equiv (z-z_m)/\Delta z$.
Standard recombination is well fit by $z_m\simeq 1200$ and $\Delta z\simeq 80$.
A sampling of the spectra, computing using the code described in
White \& Scott~(\cite{WhiSco}), are shown in Fig.~\ref{fig:zls}.
Recall that our limit on $z_m$ will depend slightly on the particular
cosmological parameters chosen, but should be very insensitive to the
details of the model for structure formation (e.g.~inflationary CDM).

A modification to the time or duration of recombination was discussed as a
solution to the apparently missing second peak in the Boomerang data
(de Bernardis et al.~\cite{Boom}; Hu~\cite{Hu00};
White, Scott \& Pierpaoli~\cite{WSP}; Peebles, Seager \& Hu~\cite{PeeSeaHu};
Tegmark \& Zaldarriaga~\cite{TegZal}).
{}From Fig.~\ref{fig:zls} a detection of significant anisotropy at
$\ell\ga 1000$ constrains the last scattering surface to lie at $z\ga 800$
for the ``concordance'' cosmology.

Our picture of the ionization history of the universe is thus as follows:
The visibility of flux shortward of Ly-$\alpha$ in high-$z$ quasars
(e.g.~Fan et al.~\cite{SDSS}) indicates that the universe is highly
ionized back to $z\simeq 6$.
The detection of anisotropy on degree scales indicates that the
universe was neutral above a redshift of $\sim 30$
(Tegmark et al.~\cite{TegZalHam}).
We have now been able to demonstrate that the universe reionized between
$z\sim 800$ and $z\sim 1600$ when the temperature was $2000-4000$K.
This is, not surprisingly, in accord with our understanding of recombination
physics (e.g.~Seager, Sasselov \& Scott~\cite{SSS}) in the `standard'
cosmology and we expect the universe was ionized at all higher redshifts.

\begin{figure}[tbh]
\leavevmode
\centerline{\epsfxsize=3.5in \epsfbox{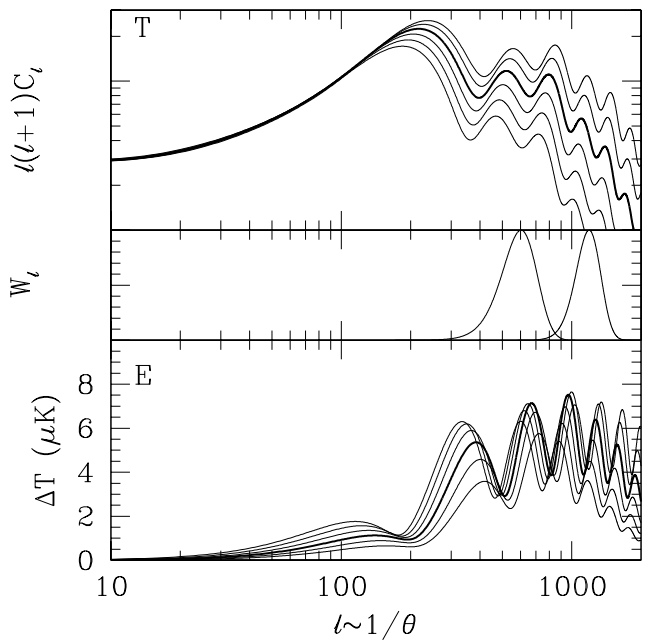}}
\caption{\footnotesize (Top panel) Temperature anisotropy spectra for the
concordance model if recombination is modeled as a Fermi function centered
at $z_m=1200$ of width $\Delta z=40$ (top curve) through $140$ (bottom curve)
in steps of $20$.  The curves have been normalized to agree at low-$\ell$.
Standard recombination is close to $\Delta z=80$ (thick line).
(Middle panel) Our approximate window functions for the CBI bandpowers,
normalized to unity at peak.
(Bottom panel) The $E-$mode polarization in the models above, with
$\Delta T=\sqrt{\ell(\ell+1)C_\ell/(2\pi)}$.
Higher polarization at $\ell\sim 300$ corresponds to larger $\Delta z$.}
\label{fig:recomb}
\end{figure}

While the above limit is on the distance to the last scattering surface, we
can also limit the thickness of the last scattering surface.
A modification to the duration of recombination was discussed as a solution
to the apparently missing second peak in the Boomerang data as described
above.
Because the power at $\ell\sim 10^3$ is ``low'' we know the duration of
recombination can't be too short.  If recombination proceeded as quickly
as the standard Saha theory (e.g.~Lang~\cite{Lan}) would predict for example,
we would expect significantly more power at $\ell\sim 10^3$ than is observed
(see Fig.~9 of Hu et al.~\cite{HSSW}).
To demonstrate the sense of this effect we have modified $\Delta z$ in our
mockup of recombination described above.
A sampling of the spectra with $\Delta z$ ranging from 40 to 140
are shown in Fig.~\ref{fig:recomb}.
As a point weighted towards $\ell\simeq 1500$ is not available, the unknown
distance to last scattering introduces considerable uncertainty in the
upper limit on $\Delta z$ using only the CBI data.
We can see however that recombination cannot be much shorter than
$\Delta z\sim 50$ --  recombination to the ground state is inhibited by
the recombination photons (Novikov \& Zel'dovich~\cite{NovZel}).

Our lower limit on $\Delta z$ nicely brings us to the next fundamental CMB
milestone -- detection of polarization.
There are extremely strong theoretical reasons to believe that the anisotropy
is polarized at a low level, since the angular dependence of Thomson
scattering is sensitive to polarization.  To date only upper limits have been
reported
(e.g.~Staggs, Gundersen \& Church~\cite{StaGunChu};
 Hedman et al.~\cite{HBGSW}),
but they remain above the theoretical predictions of popular models.
Since polarization arises from scattering, it is generated `during' last
scattering and possibly in a second, closer, scattering surface during
reionization.
The thicker the last scattering surface the stronger the polarized signal,
but also the larger the angular scale at which damping becomes effective.
Our lower limit on the duration of recombination is thus a lower limit on
the polarization of the sub-degree scale anisotropy.
This is shown in the lower panel of Fig.~\ref{fig:recomb} where the thicker
last scattering surfaces have enhanced power at $\ell\sim 300$, and a peak
polarization signal shifted to lower $\ell$ by the increased damping.

\section{Conclusions}

Our theories suggest that the CMB power spectrum consists of three regions,
separated by two physical scales (the sound horizon and the damping length).
With the detection of power on arcminute scales by the CBI experiment,
all of the major parts of the temperature anisotropy spectrum have been
observed: the low-$\ell$ plateau, the acoustic peaks and the damping region.
The results are strikingly similar to theoretical predictions made nearly two
decades ago, lending further support to a model in which the large scale
structure grew through gravitational instability from small primordial
perturbations in density (presumably laid down by inflation).

For many years theorists have been describing what we may learn from ``future''
measurements of the damping tail.  Here we have begun to implement this
program.
A number of constraints on parameterized models in the CDM family have
already been presented by Padin et al.~(\cite{CBI}) on the basis of the
data released to date.
We have presented some different and more model independent constraints,
including constraints on the matter density, the distance to last scattering
and the duration of last scattering.
Popular models based on inflationary CDM pass all of these constraints easily,
while many non-standard cosmological models fare less well.
By providing a lower limit on the thickness of the last scattering surface,
the CBI measurement implies a lower limit to the polarization of the
sub-degree scale anisotropy which is marginally lower than current upper
limits.
Further data from CBI, and other experiments, at these angular scales should
enable us to extract more of the cosmological information contained in
the damping tail.

\smallskip
{\it Acknowledgments:} 
I thank J.~Cohn, T.~Pearson, A.~Readhead and D.~Scott
for a useful comments which improved the the manuscript.
M. White was supported by a Sloan Fellowship and
the US National Science Foundation.

\end{document}